\documentclass[aps,prd,nofootinbib,showkeys]{revtex4}
\usepackage{graphicx}
\usepackage{amssymb}
\usepackage{amsmath}
\usepackage[hyperfootnotes=false]{hyperref}
\begin{document}



\title{Apparent superluminal velocities and random walk in the velocity space}





\author{Abhijit Sen}
\email{abhijit913@gmail.com}
\affiliation{Novosibirsk State University, Novosibirsk 630
090, Russia}

\author{Z.K. Silagadze}
\email{Z.K.Silagadze@inp.nsk.su}
\affiliation{ Budker Institute of
Nuclear Physics and Novosibirsk State University, Novosibirsk 630
090, Russia }

\begin{abstract}
We conjecture that the random walk and the corresponding diffusion in the 
relativistic velocity space is an adequate method for describing the 
acceleration process in relativistic jets. Considering a simple toy model, 
the main features of diffusion in the velocity space are demonstrated in both 
non-relativistic and relativistic regimes.
\end{abstract}

\maketitle

\section{Introduction}
Apparent superluminal motion has been found to occur in many astrophysical 
phenomena with the relativistic ejecta such as active galactic nuclei (quasars,
radio galaxies, blazars), gamma-ray bursts and micro-quasars \cite{1,2,3,4}.
Although these objects are significantly different in their sizes and 
properties, the common feature that makes possible the apparent superluminal 
motion is that they all support relativistic jets.

In fact, the possibility of apparent superluminal motion was predicted by Rees
\cite{5} well before the phenomenon was actually observed thanks to a new
very long baseline interferometry technique, first in the quasar 3C273, and then
in many other sources \cite{2}.

\begin{figure}   
\begin{center}
\includegraphics[scale=0.50]{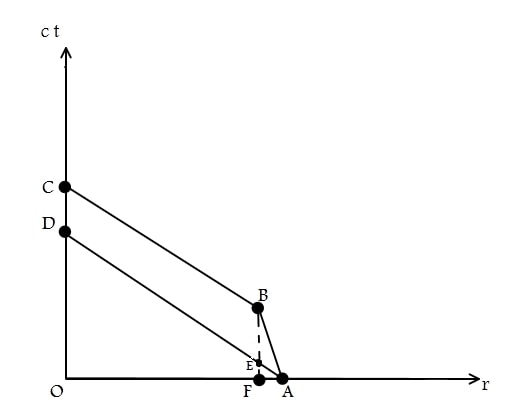}
\end{center}
\caption{Illustration of the Doppler contraction of the arrival times of 
photons.}
\label{fig1}
\end{figure}

The basic idea behind the apparent superluminal motion is rather elementary: 
this effect arises from the Doppler contraction of the arrival times of photons
due to the finite speed of light \cite{6,7}. This Doppler contraction is most 
easily explained by the space-time diagram in Fig.~\ref{fig1}.

Suppose a radio-emitting blob moves along world-line $AB$. Two consecutive
radio-emissions at $A$ and $B$ are separated by a time interval $\Delta t_e$.
Arrival times of the corresponding radio-waves at $D$ and $C$ are separated
by another time interval $\Delta t_r$. Then $BF=c \Delta t_e$ and $CD=BE=
c\Delta t_r$. If the velocity of the blob is $V$, and its radial component is
$V\cos{\theta}$, then $AF=EF=V\cos{\theta}\,\Delta t_e$. It is clear from  
Fig.~\ref{fig1} that $c\Delta t_r=BF-EF=c \Delta t_e-V\cos{\theta}\,
\Delta t_e$. Therefore $\Delta t_r=\Delta t_e (1-\beta\,\cos{\theta})$. Since 
the lateral displacement of the blob is $V\sin{\theta}\,\Delta t_e$, the 
apparent transverse velocity of the blob (divided by the light velocity $c$)  
equals to\footnote{This formula is valid only for nearby sources. The 
apparent transverse velocity of a distant radio source depends on its redshift
$z$ and is reduced by a factor of $1+z$ \cite{2,8}.}
\begin{equation}
\beta_{app}=\frac{V\sin{\theta}\,\Delta t_e}{c \Delta t_r}=\frac{\beta
\sin{\theta}}{1-\beta\cos{\theta}}.
\label{eq1}
\end{equation}
The apparent lateral motion is superluminal if
\begin{equation}
\beta>\frac{1}{\sin{\theta}+\cos{\theta}}=\frac{1}{\sqrt{2}\sin{\left 
(\frac{\pi}{4}+\theta\right )}}\ge\frac{1}{\sqrt{2}}\approx 0.71.
\label{eq2}
\end{equation}
Consequently, apparent superluminal motion is possible only if the source moves 
with relativistic speed.

Relativistic jets are produced by accreting compact objects such as neutron 
stars or black holes, although the exact mechanism how relativistic outflows 
are launched is not yet clear \cite{9}. The most promising candidate is the so 
called Blandford--Znajek mechanism \cite{10,11}, which suggests that the jets 
are driven by the rotational energy of the black hole, extracted 
electromagnetically in the form of a Poynting flux through magnetic fields 
penetrating  the event horizon of the central black hole.

However, there is still no consensus on the mechanisms that cause acceleration 
and collimation of jets. Observations show that jets are subluminal near the 
core, then they are gradually accelerated  and become relativistic at 
a distance relatively remote from the central engine (about a thousand 
Schwarzschild radii) \cite{12}.

Possible acceleration mechanisms are mostly related to the propagation of 
relativistic shock waves and turbulent structures in the jets and include 
diffusive shock acceleration (first-order Fermi acceleration), fast magnetic 
reconnection, second-order Fermi acceleration (stochastic acceleration), 
Compton rocket \cite{9,13,14,15}. There is, probably, a competition between 
different acceleration mechanisms, with stochastic acceleration playing 
a significant role \cite{13}. Due to the stochastic nature of the acceleration 
process, it was proposed to describe it as diffusion in momentum space 
\cite{16}, assuming that the isotropic and homogeneous phase-space density 
$f(p,t)$ evolves in accordance with the equation
\begin{equation}
\frac{\partial f(p,t)}{\partial t}=\frac{1}{p^2}\,\frac{\partial }{\partial p}
\left [ p^2D(p,t)\,\frac{\partial f(p,t)}{\partial p}\right ],
\label{eq3}
\end{equation}
where $D(p,t)$ is the momentum-diffussion coefficient.

Our hypothesis is that probably the better description of the stochastic 
acceleration process is given by a random walk on the relativistic velocity 
space. Even a very simple type of such random walk,  an inebriated-astronaut 
model, is surprisingly effective in producing relativistic velocities \cite{17}
in the sense that rapidity grows linearly with the number of steps, not as the 
square root of the number of steps, as expected from non-relativistic 
experience.

\section{Random walk on the non-relativistic velocity space}
We suppose that in the comoving rest frame of the particle under stochastic 
acceleration it receives isotropic kicks and after each kick its velocity 
changes by $\Delta \vec{V}$, such that $\delta V=|\Delta \vec{V}|=
\mathrm{const}$. If at the $(n-1)$-th step the velocity was $\vec{X}$ and 
after the $n$-th kick it became $\vec{V}$, then $\vec{X}+\Delta \vec{V}=
\vec{V}$ and
\begin{equation}
X=\sqrt{V^2+(\delta V)^2-2V\delta V\cos{\alpha}}\approx V-\delta V 
\cos{\alpha}+\frac{(\delta V)^2}{2V}\sin^2{\alpha},
\label{eq4}
\end{equation}
where $\alpha$ is the angle between $\Delta \vec{V}$ and $\vec{V}$. 
The probability that $\alpha$ is in the range $(\alpha,\,\alpha+\Delta\alpha)$
is $\frac{d\Omega}{4\pi}=\frac{1}{2}\,\sin{\alpha}\,d\alpha$. Therefore, if
$P(V,n)$ is the probability density at the $n$-th step\footnote{That is,
$P(V,n)\,d\vec{V}$ is the probability that the velocity at the $n$-th step 
will be in the range $(\vec{V},\,\vec{V}+d\vec{V})$.}, then we can write
\begin{equation}
P(V,n)=\int\limits_0^\pi P\left (V-\delta V \cos{\alpha}+\frac{(\delta V)^2}
{2V}\sin^2{\alpha},\, n-1\right )\frac{1}{2}\,\sin{\alpha}\,d\alpha.
\label{eq5}
\end{equation}
Now we assume that $\delta V\ll V$ and expand up to the quadratic terms in
$\delta V$:
\begin{equation}
P\left (V-\delta V \cos{\alpha}+\frac{(\delta V)^2}{2V}\sin^2{\alpha},\, n-1
\right )\approx P(V,n-1)+\frac{\partial P}{\partial V}\left (-\delta V
\cos{\alpha}+\frac{(\delta V)^2}{2V}\sin^2{\alpha}\right )+\frac{1}{2}\,
\frac{\partial^2 P}{\partial V^2}(\delta V)^2\cos^2{\alpha}.
\label{eq6}
\end{equation}
We can substitute (\ref{eq6}) into (\ref{eq5}), evaluate elementary integrals
and obtain
\begin{equation}
P(V,n)=P(V,n-1)+\frac{1}{3}\,\frac{(\delta V)^2}{V}\,\frac{\partial P}
{\partial V}
+\frac{1}{6}\,(\delta V)^2\,\frac{\partial^2 P}{\partial V^2}.
\label{eq7}
\end{equation}
If $n\gg 1$, then $P(V,n)-P(V,n-1)\approx \frac{\partial P}{\partial n}$ and
(\ref{eq7})  can be rewritten as follows
\begin{equation}
\frac{\partial P}{\partial n}=\frac{(\delta V)^2}{6V^2}\,\frac{\partial }
{\partial V}\left (V^2\frac{\partial P}{\partial V}\right)=
\frac{(\delta V)^2}{6}\,\nabla^2 P,
\label{eq8}
\end{equation}
where $\nabla^2$ is the Laplacian in the velocity space, and the last equality
follows from the fact that $P$ depends only on the magnitude $V$ of the
velocity, not on the angular variables characterizing its direction.

The equation (\ref{eq8}) is the same diffusion equation (\ref{eq3}), provided 
that the momentum is given by its non-relativistic expression $p=mV$ and the 
diffusion coefficient $D(p,t)=(\delta V)^2/6$ is a constant. We need a solution
of this diffusion equation with the initial condition $P(V,0)=f_0\delta(V)$,
which assumes that the initial velocity was zero. Here $f_0$ is the 
normalization constant to be determined from the normalization condition
$\int P(V,n)\,d\vec{V}=1$, $n\ne 0$\footnote{$P(V,0)$ cannot be used to
determine the normalization constant $f_0$, because, as can be seen from 
the explicit form of $P(V,n)$, $\lim_{n\to 0}\int P(V,n)\,d\vec{V}\ne\int\left (
\lim_{n\to 0}P(V,n)\right )d\vec{V}$.}. 

The solution is given by the well-known Euclidean heat kernel \cite{18} and has 
the form of a normalized Gaussian
\begin{equation}
P(V,n)=(4\pi)^{-3/2}\left (\frac{n\,(\delta V)^2}{6}\right )^{-3/2}
\exp{\left (-\frac{3V^2}{2n\,(\delta V)^2}\right )}.
\label{eq9}
\end{equation}
As we see,  the likelihood that the speed increase will far exceed $\sqrt {n}\,
\delta V$ is vanishingly small. As expected, the random walk efficiency for 
producing high speeds increases as $\sqrt {n}$.

For the convenience of readers, we present in the appendix the derivation of 
the Euclidean heat kernel. Although originally associated with the heat 
equation \cite{19},  physical motivation soon receded far into the background, 
and heat kernels in their general form became ubiquitous in mathematics 
\cite{20,21,22,23}.

\section{Random walk on the relativistic velocity space}
If $\delta V\ll c$, in the reference frame where initially the particle 
velocity is non-relativistic, it will remain non-relativistic for many 
consecutive kicks. Therefore, equation (\ref{eq8}) remains valid in this 
frame locally. It will be globally valid if written in the covariant form
\cite{17}
\begin{equation}
\frac{\partial P}{\partial n}=\frac{(\delta V)^2}{6}\,\nabla^2 P, \;\;\; 
\nabla^2 P=\frac{1}{\sqrt{g}}\frac{\partial}
{\partial V^\alpha}\left (\sqrt{g}\,g^{\alpha \beta}\frac{\partial P}
{\partial V^\beta}
\right ),
\label{eq10}
\end{equation}
where Einstein's summation convention is accepted, $g_{\alpha \beta}$ is the metric
tensor in the relativistic velocity space with determinant $g$, $g^{\alpha \beta}$ 
is the inverse of the metric tensor, and the expression used for the Laplacian 
in curved spaces can be found, for example, in \cite{24}.

Given two relativistic velocities, $\vec{V}$ and $\vec{V}+d\vec{V}$, the 
square of the relative velocity can be considered as a line element in 
three-dimensional velocity space \cite{25,26,27}. If $\vec{V}_1$ and 
$\vec{V}_2$ are two relativistic velocities, and $u_1$ and $u_2$ are the 
corresponding four-velocities, then it easy to see that the square of the 
relative velocity is  \cite{17,28} (to simplify expressions, $c=1$ is
assumed throughout the rest of the paper) 
\begin{equation}
[\vec{V}_1\ominus\vec{V}_2]^2=1-\frac{1}{(u_1\cdot u_2)^2}=\frac{(\vec{V}_1-
\vec{V}_2)^2-[\vec{V}_1\times\vec{V}_2]^2}{(1-\vec{V}_1\cdot\vec{V}_2)^2}.
\label{eq11}
\end{equation}
Substituting $\vec{V}$ and $\vec{V}+d\vec{V}$, insteed of  $\vec{V}_1$ and 
$\vec{V}_2$ in this formula, we get for the line element of the relativistic 
velocity frame the following expression
\begin{equation}
ds^2=\frac{(d\vec{V})^2-[\vec{V}\times d\vec{V}]^2}{(1-\vec{V}\cdot\vec{V})^2}=
\frac{dV^2}{(1-V^2)^2}+\frac{V^2}{1-V^2}\left (d\theta^2+\sin^2{\theta}\,
d\varphi^2\right ),
\label{eq12}
\end{equation}
where the last equality follows from $[\vec{V}\times d\vec{V}]^2=V^2 d\vec{V}
\cdot  d\vec{V}-(\vec{V}\cdot  d\vec{V})^2$ and $\vec{V}=V(\sin{\theta}\cos{
\varphi},\,\sin{\theta}\sin{\varphi},\,\cos{\theta})$.

It turns out that (\ref{eq12}) is the Riemannian line element corresponding to 
the Beltrami--Klein model $\mathbb{H}^3$ of three-dimensional hyperbolic 
geometry 
\cite{26,29}.

The natural parameter for Lorentz boosts is rapidity, not velocity \cite{30}. 
It is not surprising therefore that (\ref{eq11}) simplifies if, instead of the 
velocity $V$, we introduce the  rapidity $\psi$ through $V=\tanh{\psi}$:
\begin{equation}
ds^2=d\psi^2+\sinh^2{\psi}\left (d\theta^2+\sin^2{\theta}\,d\varphi^2\right ).
\label{eq13}
\end{equation}
Thus, non-zero  components of the metric tensor and its determinant are
\begin{equation}
g_{\psi\psi}=1,\;\; g_{\theta\theta}=\sinh^2{\psi},\;\; g_{\varphi\varphi}=
\sinh^2{\psi}\,\sin^2{\theta},\;\; g=\sinh^4{\psi}\,\sin^2{\theta}.
\label{eq14}
\end{equation}
Because of spherical symmetry, $P(\psi,n)$ does not depend on $\theta$ and 
$\varphi$. Therefore, from (\ref{eq10}),  we get \cite{17}
\begin{equation}
\frac{\partial P}{\partial n}=\frac{(\delta V)^2}{6}\,\frac{1}{\sinh^2{\psi}}\,
\frac{\partial}{\partial \psi}\left ( \sinh^2{\psi} \,\frac{\partial P}
{\partial \psi}\right )=\frac{(\delta V)^2}{6}\left [ \frac{\partial^2 P}
{\partial \psi^2}+2\,\frac{\cosh{\psi}}{\sinh{\psi}}\,\frac{\partial P}
{\partial \psi} \right ].
\label{eq15}
\end{equation}
To solve (\ref{eq15}) with $P(\psi,0)\sim \delta(\psi)$ initial condition, let 
us note that
\begin{equation}
\frac{\psi}{\sinh{\psi}}\,\Delta_0\,\frac{\sinh{\psi}}{\psi}=\frac{\partial^2 }
{\partial \psi^2}+2\,\frac{\cosh{\psi}}{\sinh{\psi}}\,\frac{\partial }
{\partial \psi}+1=\Delta+1,
\label{eq16}
\end{equation}
and
\begin{equation}
e^{-an}\,\frac{\partial }{\partial n}\,e^{an}=\frac{\partial }{\partial n}+a,
\label{eq17}
\end{equation}
where $a=\frac{(\delta V)^2}{6}$, and $\Delta_0$,  $\Delta$ are radial parts of
the Euclidean and hyperbolic Laplacians:
\begin{equation}
\Delta_0=\frac{1}{\psi^2}\,\frac{\partial }{\partial \psi}\,\left (\psi^2\,
\frac{\partial }{\partial \psi}\right ),\;\;\Delta=\frac{1}{\sinh^2{\psi}}\,
\frac{\partial }{\partial \psi}\,
\left (\sinh^2{\psi}\,\frac{\partial }{\partial \psi}\right ).
\label{eq18}
\end{equation}
Relations (\ref{eq16}) and (\ref{eq17}) imply that
\begin{equation}
e^{-an}\frac{\psi}{\sinh{\psi}}\left ( \frac{\partial }{\partial n}-a\Delta_0
\right )\frac{\sinh{\psi}}{\psi}e^{an}=\frac{\partial }{\partial n}-a\Delta .
\label{eq19}
\end{equation}
Therefore, if $P_0(\psi,n)$ is a solution of the Euclidean heat 
equation 
$$\left ( \frac{\partial }{\partial n}-a\Delta_0\right )P_0(\psi,n)=0$$ 
with the initial condition $P_0(\psi,0)\sim \delta(\psi)$, then 
$$P(\psi,n)=\frac{\psi}{\sinh{\psi}}e^{-an}P_0(\psi,n)$$ 
is the solution of the heat equation 
$$\left ( \frac{\partial }{\partial n}-a\Delta\right )P(\psi,n)=0$$ 
in the $\mathbb{H}^3$ hyperbolic space with the initial condition  
$P(\psi,0)\sim \delta(\psi)$.

Consequently, the required solution of  (\ref{eq15}) is
\begin{equation}
P(\psi,n)=e^{-\frac{(\delta V)^2}{6}\,n}\,\frac{\psi}{\sinh{\psi}}\,
(4\pi)^{-3/2}\left ( \frac{n\,(\delta V)^2}{6}\right )^{-3/2}\exp{\left 
(-\frac{3\psi^2}{2n\,(\delta V)^2}\right )}.
\label{eq20}
\end{equation}
This result agrees, as it should be, with the well-known expression of the heat
kernel in the $\mathbb{H}^3$ hyperbolic space \cite{22,31,32,33}. Besides, in 
the non-relativistic limit $\psi\approx V\ll 1$, $n\,(\delta V)^2\ll 1$, 
(\ref{eq20}) turns into (\ref{eq9}).

The volume element in  $\mathbb{H}^3$ is $d\vec{V}=\sqrt{g}\,d\psi d\theta 
d\varphi$. Therefore, the probability that $\psi$ will be between  $\psi$ and  
$\psi+d\psi$ is $\int\limits_0^\pi d\theta \int\limits_0^{2\pi} d\varphi \, 
\sqrt{g}\,P(\psi,n)\, d\psi=4\pi\sinh^2{\psi}\,P(\psi,n)\, d\psi$. Combined 
with (\ref{eq20}), this implies the following probability density for the 
rapidity $\psi$:
\begin{equation}
p(\psi,n)=\frac{\psi\,\sinh{\psi}}{\sqrt{4\pi}}\,\left ( \frac{n\,(\delta V)^2}
{6}\right )^{-3/2}\exp{\left (-\frac{3}{2n\,(\delta V)^2}\left [\psi^2+
\frac{n^2(\delta V)^4}{9}\right ] \right )}.
\label{eq21}
\end{equation}
In the ultra-relativistic limit $\psi\gg 1$, $\sinh{\psi}\approx \frac{1}{2}
e^\psi$ and
(\ref{eq21}) takes the form
\begin{equation}
p(\psi,n)=\frac{\psi}{\sqrt{16\pi}}\,\left ( \frac{n\,(\delta V)^2}{6}
\right )^{-3/2}\,\exp{\left (-\frac{3}{2n\,(\delta V)^2}\left [\psi-
\frac{n(\delta V)^2}{3}\right ]^2\right )}.
\label{eq22}
\end{equation}
When $n$ increases, the rapidity grows linearly with  $n$, not as  $\sqrt{n}$, 
and becomes more and more concentrated around $\psi=\frac{n(\delta V)^2}{3}$.

\section{Concluding remarks}
The observation of apparent superluminal velocities in astrophysical jets 
indicates that  particles move relativistically in these jets. To explain 
observed features of such jets,  in addition to the initial jet acceleration 
by means of the Blandford-Znajek and/or some other processes, additional 
acceleration mechanisms operating within the jets are required, with the 
stochastic acceleration playing probably a significant role  \cite{13}.  We 
conjecture that in this case random walk and the corresponding diffusion in 
the relativistic velocity space is an adequate method for describing the 
acceleration process in relativistic jets.

The formulation of a consistent theory of relativistic diffusion is 
a long-standing fascinating problem in physics \cite{34,35}. The diffusion 
equation is a parabolic differential equation and it assumes an infinite 
propagation speed for the initial data. Therefore, a relativistic diffusion 
cannot exist on the Minkowski space-time. However, as shown by 
R.~M.~Dudley \cite{36} it still makes sense at the level of its tangent 
bundle (the relativistic velocity space). We believe that the idea that 
particles could follow a Brownian motion on the relativistic velocity space 
could prove interesting in the context of astrophysics \cite{37}.

The way we got the heat kernel on $\mathbb{H}^3$ is similar (but not identical)
to the use of shift operators (intertwining operators) \cite{38,39,40} relating
radial parts of Laplacians on  $\mathbb{H}^{2n+1}$ and  $\mathbb{R}^1$.

\section*{Acknowledgments}
The work is supported by the Ministry of Education and Science of the Russian 
Federation.

\appendix
\section{Derivation of the Euclidean heat kernel}
We want to solve the heat equation 
\begin{equation}
\frac{\partial P}{\partial t}=a\,\nabla^2 P,
\label{eqa1}
\end{equation}
with the initial condition
\begin{equation}
P(\vec{r},0)=\delta(\vec{r}).
\label{eqa2}
\end{equation}
The following method of solution is adapted from \cite{18}. We take 
$P(\vec{r},t)=f(x,t)f(y,t)f(z,t)$, where $f(x,t)$ satisfies a one-dimensional
heat equation
\begin{equation}
\frac{\partial f}{\partial t}=a\,\frac{\partial^2 f}{\partial x^2},
\label{eqa3}
\end{equation}
with the initial condition
\begin{equation}
f(x,0)=\delta(x).
\label{eqa4}
\end{equation}
It is convenient to introduce a dimensionless auxiliary function 
$\tilde{f}(x,t)$, such that $f(x,t)=\frac{\partial \tilde{f}(x,t)}
{\partial x}$. If we further assume that this function satisfies the same 
one-dimensional heat equation (\ref{eqa3}), but with the different initial 
condition
\begin{equation}
\tilde{f}(x,0)=\Theta(x),
\label{eqa5}
\end{equation}
where $\Theta(x)$ is the Heaviside step function, then $f(x,t)$ will just 
satisfy (\ref{eqa3}) with the correct initial condition  (\ref{eqa4}).

According to Buckingham's Pi-theorem \cite{41}, any physical law can be 
expressed as a relationship between dimensionless quantities. From $x$, $t$ 
and $a$, we can construct only one independent dimensionless quantity 
$\tau=\frac{x}{\sqrt{at}}$. Therefore, Pi-theorem implies that
\begin{equation}
\tilde{f}(x,t)=g(\tau).
\label{eqa6}
\end{equation}
Substituting (\ref{eqa6}) into the one-dimensional heat equation, we get an 
ordinary differential equation for the unknown function $g$:
\begin{equation}
\frac{d^2 g}{d\tau^2}+\frac{\tau}{2}\,\frac{d g}{d\tau}=0.
\label{eqa7}
\end{equation}
The general solution of this equation has the form
\begin{equation}
g(\tau)=g_0+g_1\int\limits_0^\tau e^{-s^2/4} ds,
\label{eqa8}
\end{equation}
where $g_0$ and $g_1$ are some constants.

The initial condition (\ref{eqa5}) imply that $g(\infty)=1$ and $g(-\infty)=0$.
Therefore, $g_0$ and $g_1$ are determined by a linear system
\begin{equation}
g_0+g_1\int\limits_0^\infty e^{-s^2/4} ds=1,\;\;\;
g_0+g_1\int\limits_0^{-\infty} e^{-s^2/4} ds=0.
\label{eqa9}
\end{equation}
In particular,
\begin{equation}
g_1=\left [\int\limits_0^\infty e^{-s^2/4} ds-\int\limits_0^{-\infty} e^{-s^2/4} ds
\right ]^{-1}=\frac{1}{2\sqrt{\pi}}.
\label{eqa10}
\end{equation}
Then
\begin{equation}
f(x,t)=\frac{\partial }{\partial x} \left [ g_0+g_1\int\limits_0^\tau 
e^{-s^2/4} ds \right ] =\frac{1}{2\sqrt{\pi}}\, \frac{\partial \tau}{\partial x}
\, e^{-\tau^2/4}=\frac{1}{\sqrt{4\pi a t}}\exp{\left (-\frac{x^2}{4at}\right )},
\label{eqa11}
\end{equation}
and 
\begin{equation}
P(\vec{r},t)=\left (4\pi a t\right )^{-3/2}\exp{\left (-\frac{r^2}{4at}\right )}.
\label{eqa12}
\end{equation}

\end{document}